# Colossal Magnetoresistance without Mixed Valence in a Layered Phosphide Crystal

*Zhi-Cheng Wang, Jared D. Rogers, Xiaohan Yao, Renee Nichols, Kemal Atay, Bochao Xu, Jacob Franklin, Ilya Sochnikov, Philip J. Ryan, Daniel Haskel, Fazel Tafti\**

Dr. Zhi-Cheng Wang, Jared D. Rogers, Xiaohan Yao, Renee Nichols, Kemal Atay, Prof. Fazel Tafti
Departments of Physics, Boston College, 140 Commonwealth Avenue, Chestnut Hill, MA 02467, USA
E-mail: fazel.tafti@bc.edu
Bochao Xu, Jacob Franklin
Physics Department, University of Connecticut, Storrs, CT USA, 06269
Prof. Ilya Sochnikov
Physics Department, University of Connecticut, Storrs, CT USA, 06269
Institute of Material Science, University of Connecticut, Storrs, CT USA, 06269
Dr. Philip J. Ryan
Advanced Photon Source, Argonne National Laboratory, Argonne IL 60439, USA
School of Physical Sciences, Dublin City University, Dublin 9, Ireland
Dr. Daniel Haskel
Advanced Photon Source, Argonne National Laboratory, Argonne IL 60439, USA

**Abstract:** Materials with strong magnetoresistive responses are the backbone of spintronic technology, magnetic sensors, and hard drives. Among them, manganese oxides with a mixed valence and a cubic perovskite structure stand out due to their colossal magnetoresistance (CMR). A double exchange interaction underlies the CMR in manganates, whereby charge transport is enhanced when the spins on neighboring $Mn^{3+}$ and $Mn^{4+}$ ions are parallel. Prior efforts to find different materials or mechanisms for CMR resulted in a much smaller effect. Here we show an enormous CMR at low temperatures in $EuCd_2P_2$ without manganese, oxygen, mixed valence, or cubic perovskite structure. $EuCd_2P_2$ has a layered trigonal lattice and exhibits antiferromagnetic ordering at 11 K. The magnitude of CMR ($10^4$ percent) in as-grown crystals of $EuCd_2P_2$ rivals the magnitude in optimized thin films of manganates. Our magnetization, transport, and synchrotron X-ray data suggest that strong magnetic fluctuations are responsible for this phenomenon. The realization of CMR at low temperatures without heterovalency leads to a new regime for materials and technologies related to antiferromagnetic spintronics.

Colossal magnetoresistance (CMR) has been a subject of intense research due to its central place in the physics of correlated electron systems as well as its relevance to magnetic memory and sensing technologies.[1–3] The accepted paradigm of CMR is based on the manganate perovskite materials where a mixed valence of $Mn^{3+}/Mn^{4+}$ mediates a ferromagnetic double-exchange (DE) interaction and a structural Jahn-Teller (JT) distortion, which cooperatively lead to a phase transition from paramagnetic (PM) insulator to ferromagnetic (FM) metal.[4–9] As a result, the electrical resistivity shows a peak near the Curie temperature ($T_C$) which rapidly drops in response to an external magnetic field, leading to a large negative magnetoresistance known as CMR.[4,10] It has remained a challenge in materials science to deviate from this paradigm and produce a sizable CMR either near an antiferromagnetic (AFM) transition or in materials without manganese, DE interaction, and JT distortion.[11–14] Overcoming this challenge is motivated by a surge of interest in the AFM spintronic and quantum information technologies that call for new materials and mechanisms of CMR based on antiferromagnetic (AFM) ordering at lower temperatures.[15–18]

In this communication, we report the striking observation of an enormous CMR in $EuCd_2P_2$, a material which is devoid of all traditional components of CMR. $EuCd_2P_2$ does not have manganese, oxygen, a mixed valence, a DE interaction, a perovskite structure, or a JT distortion. It has an AFM order at low temperature ($T_N = 11$ K), instead of the FM order at high temperature as seen in manganates. It has a trigonal unit cell with alternating layers of edge-shared $EuP_6$ octahedra and $CdP_4$ tetrahedra, different from the cubic lattice of manganates. We will show that strong magnetic fluctuations within the layered structure of this material provide a new mechanism for CMR that is aligned with the current progress in AFM spintronics.

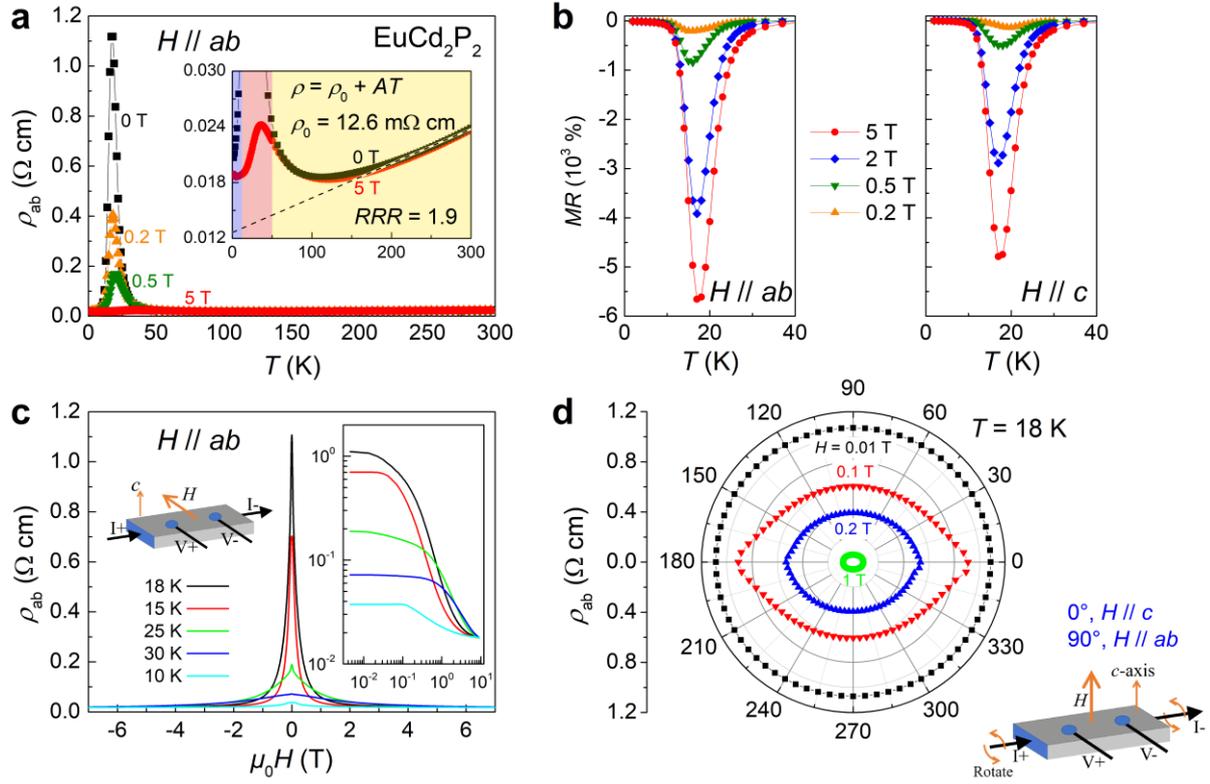

**Figure 1.** (a) Temperature dependence of resistivity at several fields. The dashed line in the inset is a linear fit to extract the residual resistivity from the zero-field data (black squares). (b) Magnetoresistance in a narrow temperature range in both in-plane and out-of-plane field directions. (c) Field dependence of resistivity at several temperatures. Inset shows two orders of magnitude drop in resistivity in less than 4 T. (d) Angle dependence of resistivity at several fields, showing a mild anisotropy. The 0 and 90 degrees correspond to $H \| c$ and $H \| ab$, respectively.

A typical in-plane resistivity curve $\rho_{ab}(T)$ from a EuCd$_2$P$_2$ single crystal reveals a large peak at 18 K in zero magnetic field (**Figure 1**a). The peak drops by 6-fold at only 0.5 T and by 46-fold at 5 T. Inset of Figure 1a shows that $\rho(T)$ fits to a bad metal behavior[19] where the high-temperature resistivity and even the residual resistivity ($\rho_0$ = 12.6 m$\Omega$ cm) are well above the Ioffe-Regel limit (1 m$\Omega$ cm). In fact, the resistivity barely changes with temperature and the residual resistivity ratio (RRR = $R_{300K}/R_0$) is only 1.9. We highlight three temperature regimes in the inset of Figure 1a: an initial poor metallic regime at high temperatures in yellow, an intermediate regime in red with CMR due to magnetic fluctuations (discussed below), and a blue region at low temperatures where the AFM order sets in and CMR disappears. Note that unlike the manganates,[10] the resistivity of EuCd$_2$P$_2$ remains much larger than 1 mΩ cm even at the lowest temperatures.

We define magnetoresistance as MR = 100%×($R_H$ − $R_0$)/$R_H$ and plot it as a function of temperature at a few representative fields in Figure 1b. The magnitude of MR exceeds −10$^3$ % in less than 1 T regardless of the field direction. For comparison, MR in single crystals of La$_{0.75}$Ca$_{0.25}$MnO$_3$, the archetypal manganate material, is only −25% at 1 T and −400% at 4 T with the same definition of magnetoresistance.[10] It reaches −10$^3$ % in epitaxially grown thin films of La$_{0.67}$Ca$_{0.33}$MnO$_3$ or La$_{0.60}$Y$_{0.07}$Ca$_{0.33}$MnO$_x$, and can be as high as −10$^{4-5}$ % after optimizing the oxygen content.[20–22] It is remarkable that as-grown crystals of EuCd$_2$P$_2$, without any material optimization, can exhibit a truly colossal effect. This will be later enhanced when we examine the effect of current direction on CMR and find a staggering −10$^4$ % MR when the current is out-of-plane instead of in-plane ($\rho_c$ instead of $\rho_{ab}$).

Both the temperature and field dependence of CMR are extremely sharp in EuCd$_2$P$_2$, making it a good material for the low-temperature magnetic sensing and read/write devices.[16] Figure 1c shows that the field dependence of resistivity has a distinct Lorentzian peak shape with a narrow full-width at half-maximum (FWHM) of 0.38 T at 18 K (see also Figure S1). Note that CMR maximizes at 18 K which is 1.5$T_N$. It becomes negligible below $T_N$ and above 5$T_N$, i.e. outside the temperature regime of magnetic fluctuations (insets of Figure 1a,c).

The CMR behavior in EuCd$_2$P$_2$ is nearly independent of the field direction, as seen in the 360° scan of the resistivity at 18 K in Figure 1d. There is no discernible angular dependence in $\rho$(18 K) at a small field of 100 Oe, and the maximum anisotropy is only a factor of 1.5 at $H$ = 0.1 T. Furthermore, the CMR does not change by changing the in-plane current direction (Figure S2); however, it increases by one order of magnitude when the current direction is changed from in-plane to out-of-plane as will be seen later.

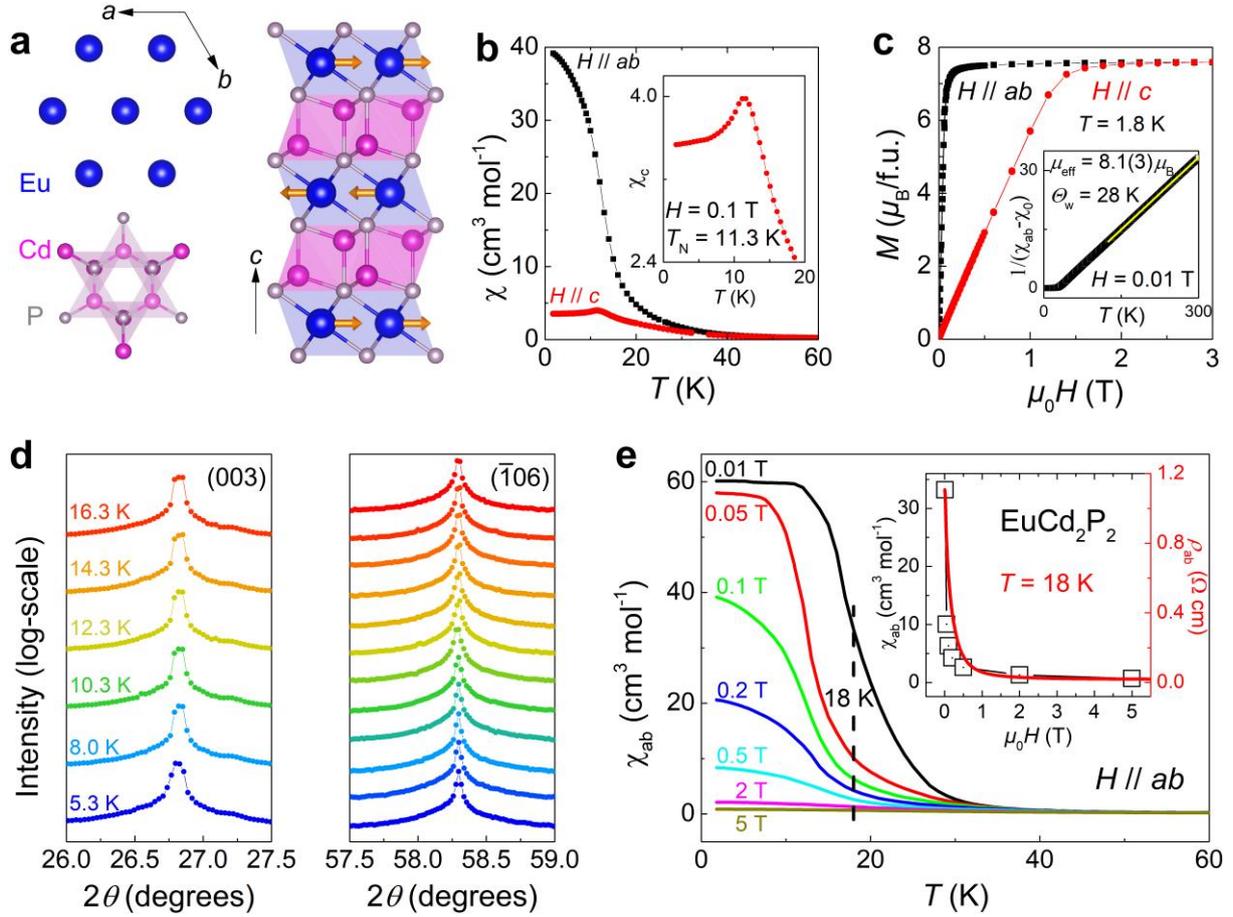

**Figure 2.** (a) Trigonal lattice of $EuCd_2P_2$ made of triangular layers of Eu and Cd with octahedral and tetrahedral coordinations, respectively. A magnetic unit cell is illustrated with A-type AFM order. (b) Magnetic susceptibility as a function of temperature in both the in-plane (black) and out-of-plane (red) fields. The inset determines $T_N$ = 11.3(2) K. (c) Magnetization as a function of both in-plane (black) and out-of-plane (red) fields. The inset determines both the effective moment $\mu_{eff}$ and Weiss temperature $\Theta_W$ from a Curie-Weiss analysis (see also Figure S5). (d) High-resolution synchrotron diffraction data are collected for two Bragg peaks, (003) and ($\bar{1}$06), at different temperatures. There is no indication of a lattice distortion at $T_N$. (e) Magnetic susceptibility as a function of temperature at several fields. The inset compares the field dependence of $\chi(T = 18\text{ K})$ and $\rho(T = 18\text{ K})$.

To investigate the underlying mechanism of CMR in $EuCd_2P_2$, we measured magnetization as a function of temperature and field in both the in-plane and out-of-plane directions. It is helpful to examine the crystal structure before discussing the magnetic data. **Figure 2**a shows the trigonal lattice of $EuCd_2P_2$ in space group $P\bar{3}m1$ with alternating Eu and Cd layers (see X-ray analysis in Figure S3. The individual layers are triangular networks of either edge-shared $EuP_6$ octahedra or CdP$_4$ tetrahedra. We show one magnetic unit cell in Figure 2a where the order is FM within the Eu layers but with alternating direction (AFM) between the layers. This A-type AFM order has been previously reported in crystals of EuCd$_2$As$_2$ and EuCd$_2$Sb$_2$,[23–27] and is consistent with the magnetic susceptibility data in Figure 2b that shows a FM order when $H \| ab$, but an AFM order when $H \| c$. The finite residual $\chi_c(T)$ near zero temperature indicates a small out-of-plane spin canting superposed on the A-type AFM order. At low temperatures, the in-plane susceptibility (black data) is 10 times larger than the out-of-plane one (red data), suggesting a strong magnetocrystalline anisotropy. This is confirmed in Figure 2c where the saturation field for the in-plane $M(H)$ curve (0.16 T) is 10 times smaller than the out-of-plane $M(H)$ curve (1.6 T). Thus, EuCd$_2$P$_2$ has a significant easy-plane anisotropy consistent with its layered structure. Inset of Figure 2b indicates $T_N$ = 11.3(2) K from the peak in $\chi_c(T)$, in agreement with a peak in the zero-field heat capacity (Figure S4).

A combination of magnetization and X-ray data confirm that the CMR in EuCd$_2$P$_2$ is unrelated to either heterovalency or JT distortions, unlike in manganates. Figure 2c establishes a fixed Eu$^{2+}$ oxidation state ($4f^7$ configuration), since both the saturated magnetization at low-$T$ (7.6(6) $\mu_B$) and the effective moment from a Curie-Weiss fit at high-$T$ (8.1(3) $\mu_B$) are consistent with the expected values for Eu$^{2+}$ (7 and 8 $\mu_B$, respectively). Note that the expected effective moment for Eu$^{3+}$ ($4f^6$) is zero, hence we rule out a mixed valence of Eu$^{2+}$/Eu$^{3+}$ unambiguously. In addition to the magnetization data, an analysis of the X-ray absorption spectroscopy in Figure S6 directly confirms the Eu$^{2+}$ oxidation state without a mixed valence. Next, we used synchrotron X-rays to trace the temperature dependence of two representative diffraction peaks, (003) and ($\bar{1}$06), in Figure 2d. There is no abrupt shift or splitting of either peak as the temperature is varied through $T_N$ = 11.3 K, hence we rule out a structural distortion (see also Figure S7).

Considering the absence of both heterovalency and lattice distortions in EuCd$_2$P$_2$, the mechanism of CMR in this material must be different from that of the manganates.[2,8,9] As noted earlier, the CMR in EuCd$_2$P$_2$ is maximum at 1.5$T_N$ and disappears either below $T_N$ or above 5$T_N$. This is the first indication that the magnetic fluctuations above $T_N$ are related to CMR. The rapid suppression of χ(T) with field in Figure 2e confirms the presence of strong magnetic fluctuations that are suppressed by field. We make a cut through the χ(T) curves at T = 18 K, where CMR is maximum, and compare the field dependence of χ(18 K) and ρ(18 K) in the inset of Figure 2e. The parallel behavior between χ(18 K) and ρ(18 K) suggests that the suppression of magnetic fluctuations with field is responsible for the CMR in this material. Such spin fluctuations are consistent with recent theoretical work on EuCd$_2$As$_2$ which has a similar layered structure as EuCd$_2$P$_2$ where the spins are confined within the 2D Eu layers (Figure 2a). [28]

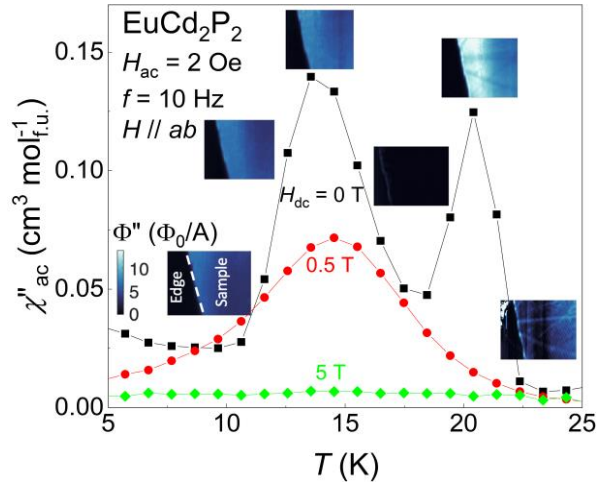

**Figure 3.** The imaginary component of the AC susceptibility data from a bulk sample is obtained at zero-field (black) as well as 0.5 and 5 T (red and green). The peaks at 13 and 20 K are due the AFM order and fluctuations, respectively. The latter peak is suppressed with a small magnetic field similar to the CMR behavior. Zero-field scanning SQUID images with the size of 220×205 μm$^2$ are compared to the bulk data at crests and troughs. The streaks in the image are due to surface roughness.

A more direct evidence for the spin fluctuations as the mechanism of CMR in EuCd$_2$P$_2$ comes from the AC susceptibility data in **Figure 3**. Two peaks are observed in the imaginary component $\chi''_{ac}(T)$. The lower temperature peak coincides with the AFM peak in the DC susceptibility (Fig. 2b), but the higher temperature peak does not have any counterpart in the DC data, i.e. it can be regarded as a direct evidence of magnetic fluctuations. Remarkably, a small DC field of 0.5 T is enough to suppress this peak, similar to the rapid suppression of the resistivity peak at 18 K (Figure 1a). Therefore, a direct link is established between the peak from magnetic fluctuations in $\chi''_{ac}(T)$ and the peak in $\rho(T)$. We also provide scanning SQUID microscopy[29] images in Figure 3 to confirm the bulk measurements. The bright and dark images correspond to the peaks and valleys in the bulk $\chi''_{ac}(T)$ data. The images highlight spatial uniformity of the AC susceptibility and confirm the intrinsic origin of the magnetic fluctuations. Although our data are consistent with spin fluctuations as the source of CMR in $EuCd_2P_2$, other mechanisms such as band structure reconstruction and topological effects are also relevant to this material.[24,32]

The A-type AFM order in $EuCd_2P_2$ resembles a magnetic tunnel junction (MTJ)[30,31] as illustrated in the inset of **Figure 4**a. A consequence of such a magnetic structure is that CMR reaches $-10^4$ % when the electric current is out of plane ($J\|c \rightarrow \rho_c$) and becomes even larger than the $-10^3$ % effect with in-plane current ($J\|ab \rightarrow \rho_{ab}$). At 5 T MR is $-28,000\%$ when $J\|c$ and $-7,600\%$ when $J\|ab$, in the same sample (inset of Figure 4a). It is noteworthy that the enormous $-10^4$ % CMR in a crystal of $EuCd_2P_2$, grown inside a hot crucible without any optimization, is three times larger than in optimized samples of $La_{0.60}Y_{0.07}Ca_{0.33}MnO_x$.[21] We point out that the magnitude of CMR is nearly independent of the field direction regardless of whether the electric current is in-plane (Figure 1d) or out-of-plane (Figures S8 and S9). The relation between CMR and magnetic fluctuations is highlighted in Figure 4b that shows the zero-field $\rho_c$ is ten times larger than $\rho_{ab}$ in the region of magnetic fluctuations (the red area). In contrast, $\rho_c$ is only two times larger than $\rho_{ab}$ in both the bad metal regime at high-$T$ (yellow) and the ordered regime at low-$T$ (blue).

Additional insight into the physics of $EuCd_2P_2$ comes from a comparison between three $EuCd_2X_2$ compounds with X = P, As, and Sb. One expects that the smaller spatial extension of the $p$-orbitals, from Sb (5$p$) to As (4$p$) and P (3$p$), reduces the coupling between the Eu and Cd layers and enhances the magnetic fluctuations, leading to a larger CMR. This is confirmed in Figure 4c where CMR increases from −10% in EuCd$_2$Sb$_2$ to −10$^2$ % in EuCd$_2$As$_2$, and −10$^3$ % in EuCd$_2$P$_2$ (with $J \| ab$). Simultaneously the exchange correlations are expected to grow from X = Sb to As and P, leading to an increase in $T_N$ from 7 K to 9 K and 11 K, respectively.[23–25] The buildup of correlations is evident in Figure 4d where both the violation of the Ioffe-Regel limit ($\rho > 1$ mΩ cm) in the bad metal regime (inset) and the resistivity peak near $T_N$ are dramatically enhanced by replacing Sb with As and P.

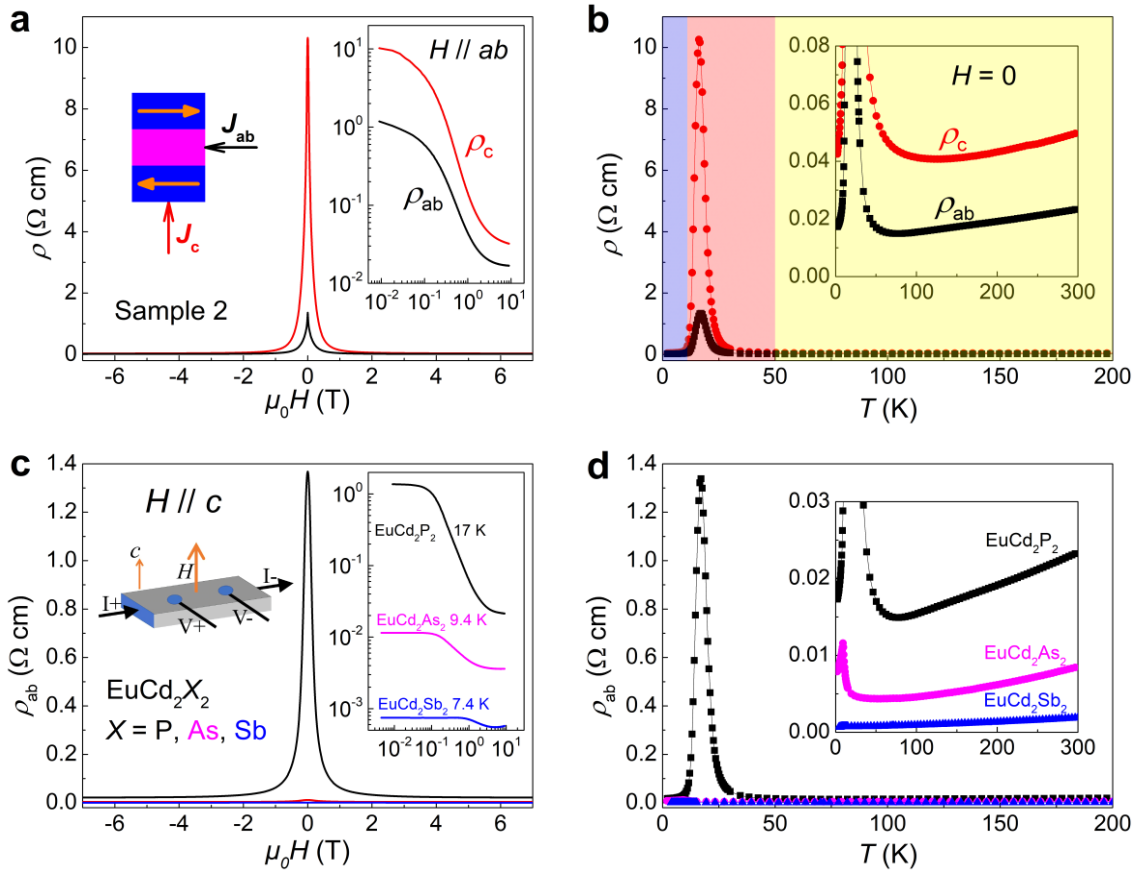

**Figure 4.** (a) CMR in EuCd$_2$P$_2$ is one order of magnitude larger when the electric current is out-of-plane compared to in-plane. The left inset shows that each unit cell of EuCd$_2$P$_2$ resembles a MTJ. The right inset shows changes of $\rho_c$ and $\rho_{ab}$ in the log-scale. The data in this figure are from Sample 2, but the data in Figures 1 and 2 are from Sample 1. (b) The magnitude of $\rho_{ab}$ and $\rho_c$ are comparable below $T_N$ (blue region) and in the bad metal regime (yellow), but they are an order of magnitude apart in the region of magnetic fluctuations (red). (c) Field dependence of resistivity in EuCd$_2$P$_2$ (black) compared to EuCd$_2$As$_2$ (magenta), and EuCd$_2$Sb$_2$ (blue). (d) The colossal resistivity peak at low temperatures in EuCd$_2$P$_2$ makes the other two materials invisible. Inset: the room temperature resistivity increases from 2 mΩ cm in EuCd$_2$Sb$_2$ to 8 mΩ cm in EuCd$_2$As$_2$ and 23 mΩ cm in EuCd$_2$P$_2$.

To summarize, we have shown that EuCd$_2$P$_2$ exhibits a CMR that reached −10$^4$ % due to strong magnetic fluctuations despite the absence of mixed valence and lattice distortions. Prior attempts to establish CMR without mixed valence has led to a much smaller effect, between −20 to −250%, e.g. in TlMn$_2$O$_7$ (pyrochlore),[11] FeCr$_2$S$_4$ (spinel),[12] and (Eu,Yb)$_{14}$MnSb$_{11}$ (Zintl compounds).[13,14] The CMR values of these materials are summarized in Table 1 and compared to EuCd$_2$P$_2$. Recently, a larger CMR of about −300% has been found in the Zintl compound EuIn$_2$P$_2$, which is also a layered material with hexagonal lattice ($P6_3/mmc$).[32] We believe that in this material, similar to EuCd$_2$P$_2$, CMR results from the layered structure and magnetic fluctuations although band structure and topological effects are also plausible mechanisms. However, due to In-In bonds within the layers, the magnetic anisotropy and the bad metal behavior are less prominent in EuIn$_2$P$_2$, as evident from a slow saturation of the in-plane magnetization and a resistivity of 2 mΩ cm at room temperature.[32] The absence of direct Cd-Cd bonds in the structure of EuCd$_2$P$_2$ seems to be beneficial to the CMR.

**Table 1.** Summary on the CMR values in several materials including oxides, chalcogenides, and pnictides.

| Reference | Material | CMR value | Field |
|---|---|---|---|
| 10 | La$_{0.75}$Ca$_{0.25}$MnO$_3$ | −25%, −400% | 1 T, 4 T |
| 11 | TlMn$_2$O$_7$ | −200~250% | 8 T |
| 12 | FeCr$_2$S$_4$ | −25% | 6 T |
| 13 | Eu$_{14}$MnSb$_{11}$ | −66% | 5 T |
| 14 | Yb$_{14}$MnSb$_{11}$ | −20% | 5.5 T |
| 32 | EuIn$_2$P$_2$ | −298% | 5 T |
| 33 | EuIn$_2$As$_2$ | −143% | 5 T |
| This work | EuCd$_2$P$_2$ | −6×10$^3$% for $I /\!/ ab$, −2.8×10$^4$% for $I /\!/ c$ | 5 T |

Due to its layered structure, intermetallic composition, and stability in air, EuCd$_2$P$_2$ is suitable for the fabrication of low temperature magnetic sensing and read/write devices. For example, lithographic techniques can be used to produce microscale spintronic devices from the single crystals. Since CMR is large regardless of the field direction (Figure 1d and S9), $EuCd_2P_2$ is convenient to work with, as the field does not need to be orientated along a specific crystallographic direction for the desired effect. Epitaxial techniques can be used to fabricate thin films and heterostructures from this layered compound for AFM spin torque and spin valve devices.[17,18] It is possible to even change the magnetic state of the material by altering the flux growth condition, as recently reported for $EuCd_2As_2$,[25] and to replace both Eu and Cd with other rare-earths and transition metals.[34] Future efforts in chemical doping, electrical biasing, and mechanical straining will enable tuning of the magnetic fluctuations, hence controlling the temperature/field regime of CMR in $EuCd_2P_2$ and its derivatives.

**Experimental Section**

*Crystal growth*: Single crystals of $EuCd_2P_2$ were grown in Sn flux, by using sublimed ingots of europium (Alfa Aesar, 99.9%), cadmium tear drops (Alfa Aesar, 99.95%), red amorphous phosphorus powder (Alfa Aesar, 98.9%), and tin shots (Alfa Aesar, 99.999%) as the starting materials. Eu ingots were cut into small pieces and mixed with other elements with a mole ratio Eu: Cd: P: Sn = 1: 2: 2: 20. The mixture was then loaded into an alumina crucible inside an evacuated quartz ampule and slowly heated to 950 °C, held for 36 h, cooled to 550 °C at 3 °C/h, and finally centrifuged to remove the excess flux.

*Transport, Heat capacity, and Magnetization Measurements*: The electrical resistivity was measured with a standard four-probe technique using a Quantum Design Physical Property Measurement System (PPMS) Dynacool with a high-resolution rotator option. The heat capacity was measured using the PPMS with a relaxation time method on a carefully polished sample. A flat crystal (1.2 mg) was adopted to measure DC magnetization using a Quantum Design Magnetic Property Measurement System (MPMS3).

*X-ray Diffraction*: Synchrotron X-ray diffraction measurements were performed at the Advanced Photon Source at beamline 6-ID B using a PSI diffractometer. The single crystal sample was cooled with a 4 K ARS cryostat refrigerator. The diffraction matrix of the sample was aligned with the (003) and ($\bar{1}$06) reflections with an X-ray energy of 11.712 keV. Temperature dependence reflection angular position was monitored by realigning the sample at each temperature before the data was taken. The crystal structures of $EuCd_2P_2$ and $SrCd_2P_2$ were refined using the powder X-ray diffraction data obtained in house. A Bruker D8 ECO instrument was used with 40 keV copper source and a 1D LYNXEYE XE detector and the FullProf suite was used for the structural refinements.

*X-ray Absorption*: X-ray absorption data were collected at beamline 4-ID-D of the Advanced Photon Source at Argonne National Laboratory. The measurements were done on a finely ground powder sample with total thickness optimized for transmission geometry. The samples were cooled in $^4$He vapor using the variable temperature insert of a superconducting magnet. Data were collected across the magnetic ordering temperature, both in zero field and $H$=2 T applied field.

**Supporting Information**

See below.

**Acknowledgements**

We thank M. Frith, Z.-X Shen, and T. Devereaux for helpful discussions. The work at Boston College was funded by the National Science Foundation under Award No. NSF/DMR-1708929. The work performed at the Advanced Photon Source was supported by the U.S. Department of Energy, Office of Science, and Office of Basic Energy Sciences under Contract No. DE-AC02-06CH11357. I. S. thanks the US DOD for partial support.

**Conflict of Interest**

The authors declare no conflict of interest.

# Supplementary Information: Colossal magnetoresistance due to strong magnetic fluctuations in a layered phosphide crystal

*Zhi-Cheng Wang, Jared D. Rogers, Xiaohan Yao, Renee Nichols, Kemal Atay, Bochao Xu, Jacob Franklin, Ilya Sochnikov, Philip J. Ryan, Daniel Haskel, Fazel Tafti*∗

## A. CMR peak shape

In the main text (Figure 1c), we showed the field dependence of CMR in $EuCd_2P_2$ with $H\|ab$. For completeness, here we show the field dependence with $H\|c$ in Figure S1. The peak shape is nearly the same whether $H\|c$ (Figure S1) or $H\|ab$ (main text, Figure 1c). We fit the CMR peak at 18 K to a Lorentzian function in the inset of Figure S1 and report the full-width at half-maximum FWHM = 0.38 T.

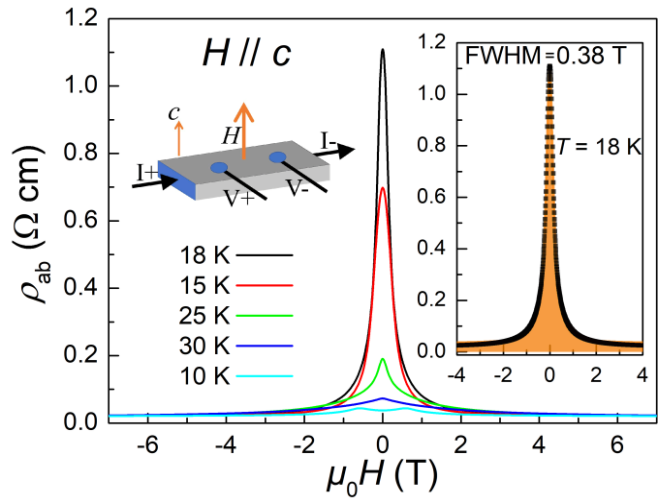

**Figure S1.** Field dependence of resistivity with $H\|c$ in $EuCd_2P_2$. The inset shows a Lorentzian peak (orange curve) with FWHM = 0.38 T.

## B. In-plane resistivity anisotropy

We characterized the in-plane resistivity anisotropy in a polished crystal of $EuCd_2P_2$, by applying out-of-plane field ($H\|c$) and in-plane current with two orthogonal directions ($H\|ab$ and $H\|a$). The results do not show any discernible anisotropy in MR. The magnitude of the resistivity is slightly different between the two measurements, but that could be due to the uncertainty in contact geometries. When we plot MR in percentage (the two bottom panels), we observe identical values. The data were obtained from Sample 2.

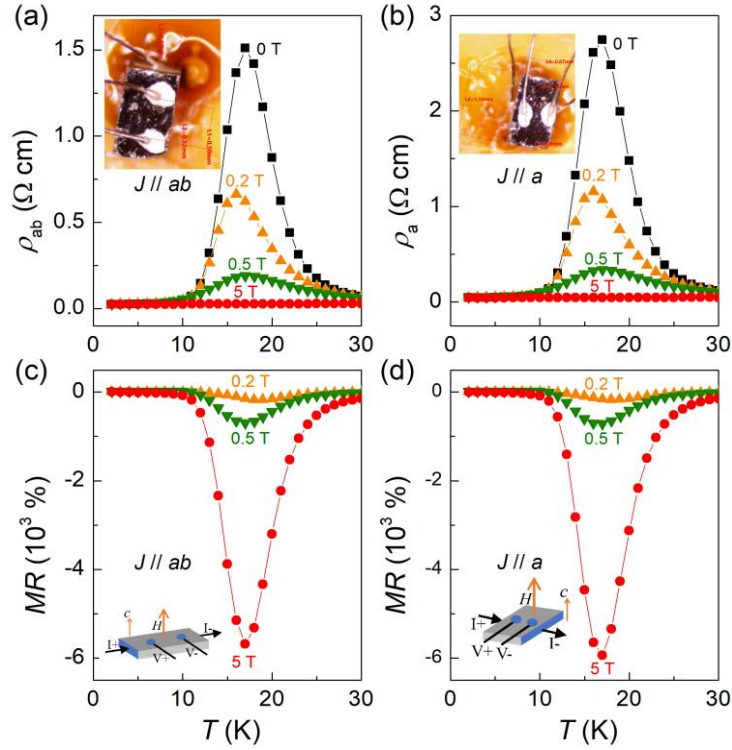

**Figure S2.** In-plane anisotropy is investigated on Sample 2. The magnitude of MR does not change by rotating the current direction. (a) and (b) show the temperature dependence of in-plane resistivity with current in two orthogonal directions ($J\|ab$ and $J\|a$). (c) and (d) show MR calculated from the resistivity data in (a) and (b). The photos of real contacts and corresponding illustrations are shown as the insets.

## C. Structural analysis

Figure S3 shows the crystallographic refinement of both EuCd$_2$P$_2$ and its lattice model SrCd$_2$P$_2$ in the trigonal space group $P\bar{3}m1$ (#164). Powder diffraction data were taken on a polycrystalline specimen. The peak positions from single crystals (ground to powder) are the same as in the polycrystalline samples. Sometimes the X-ray patterns are slightly different between the polycrystalline and single crystal specimens due to minor differences in stoichiometry or defects, but we do not find such issues here. A summary of all refinement parameters is provided in Table S1 and Table S2.

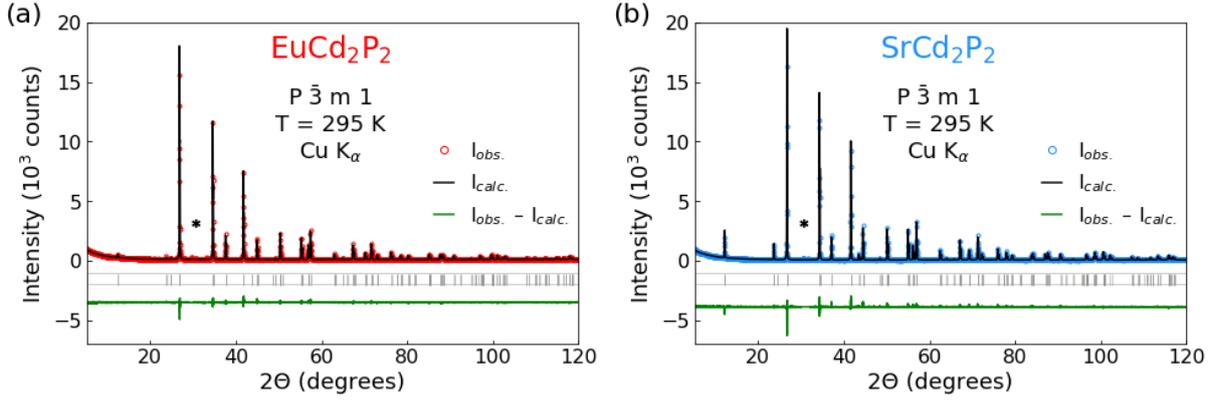

**Figure S3.** Rietveld fit (black) on the powder X-ray diffraction data (red) from EuCd$_2$P$_2$ and SrCd$_2$P$_2$ (blue). The *hkl* indices in space group #164 are marked with black tics and the fitting residual curves are shown with green lines. The asterisks mark a few tiny reflections near 30 degrees due to Eu$_3$(PO$_4$)$_2$ and Sr$_{10}$(PO$_4$)$_6$O impurities in the powder specimen used for the refinements.

### D. Magnetic heat capacity

Figure S4a shows the heat capacity of EuCd$_2$P$_2$ and its lattice counterpart SrCd$_2$P$_2$ used for phonon subtraction. The peak at 11.1(4) K indicates $T_N$ which agrees with the peak at 11.3(2) K in $\chi_c$ (main Figure 2b). We compared the AFM transition between EuCd$_2$P$_2$, EuCd$_2$As$_2$, and EuCd$_2$Sb$_2$ in Figure S4a with respective $T_N$ = 7.1, 9.2, and 11.1 K. A wider transition in EuCd$_2$P$_2$, compared to its sister compounds, shows that the magnetic fluctuations are stronger when the *p*-orbitals are less extended (3*p* in P, compared to 4*p* and 5*p* in As and Sb).

Table S1: Unit cell dimensions and refinement parameters are listed for both EuCd$_2$P$_2$ and SrCd$_2$P$_2$ from the Rietveld refinements in the space group $P\bar{3}m1$ (Figure S3).

| Material | EuCd$_2$P$_2$ | SrCd$_2$P$_2$ |
|---|---|---|
| Mass (g/mol) | 438.734 | 374.389 |
| *a* (Å) | 4.3248(2) | 4.3376(1) |
| *c* (Å) | 7.1771(7) | 7.2707(6) |
| *V* (Å$^3$) | 116.26 | 118.47 |
| Z | 1 | 1 |
| *D* (g/cm$^3$) | 6.27 | 5.25 |
| $R_p$ | 6.74 | 8.41 |
| $R_{exp}$ | 6.88 | 7.05 |
| $\chi^2$ | 1.66 | 2.45 |

Table S2: Wyckoff sites and atomic coordinates in both EuCd$_2$P$_2$ and SrCd$_2$P$_2$. The isotropic Debye-Waller factors ($B_{iso}$) are less than 1.0 Å$^2$ for all atoms. All sites are fully occupied.

| | | EuCd$_2$P$_2$ / SrCd$_2$P$_2$ | | |
|---|---|---|---|---|
| Atom | site | x | y | z |
| Eu/Sr | 1a | 0 | 0 | 0 |
| Cd | 2d | 0.33333 | 0.66667 | 0.6357(1) / 0.6345(7) |
| P | 2d | 0.33333 | 0.66667 | 0.2484(1) / 0.2496(1) |

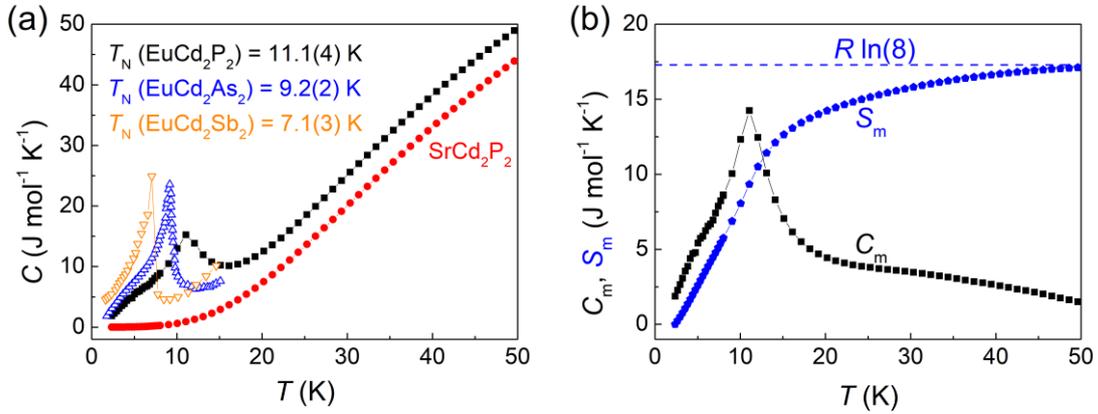

**Figure S4.** (a) The heat capacity of EuCd$_2$X$_2$ with X = P, As, and Sb as a function of temperature. Data for X = As and Sb (empty symbols) are taken from other references.[1,2] (b) The magnetic heat capacity $C_m$ (black) and entropy $S_m$ (blue) plotted as a function of temperature. The dashed line marks the spin entropy expected for Eu$^{2+}$ ions.

We plot the magnetic heat capacity $C_m = C_{EuCd2P2} - C_{SrCd2P2}$ and the magnetic entropy $S_m = \int \frac{C_m}{T} dT$ in Figure S4b. To compensate for the different molecular masses between the two compounds and to correct for different sound velocities, we multiplied the SrCd$_2$P$_2$ data by a factor of $1.08 = \sqrt{M_{EuCd2P2}/M_{SrCd2P2}}$ before subtracting it from the EuCd$_2$P$_2$ data. The magnetic entropy in Figure S4b reaches the expected value of $Rln(2S + 1)$ for Eu$^{2+}$ with S = 7/2 as indicated by the blue dashed line.

### E. Curie-Weiss analysis

We present the Curie-Weiss (CW) analysis at a small field of 0.01 T in both the in-plane ($\chi_{ab}$)

and out-of-plane ($\chi_c$) field directions in Figure S5a,b. The right *y*-axes correspond to $1/(\chi - \chi_0)$ where $\chi_0$ is a small paramagnetic background. The red line shows our CW fit according to $\chi - \chi_0 = C/(T - \Theta_W)$. The Weiss temperature $\Theta_W$ and the effective moment $\mu_{eff}$ are comparable for both directions, but the susceptibility at low temperatures is 10 times larger with in-plane field (*H*∥*ab*) as compared to out-of-plane field (*H*∥*c*).

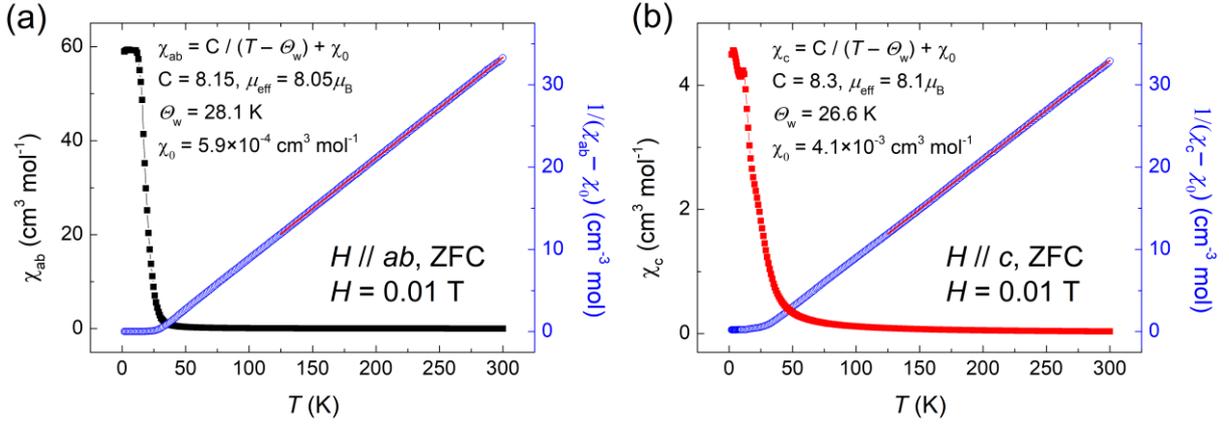

**Figure S5.** (a) CW analysis on the magnetic susceptibility data with the in-plane field. (b) CW analysis for the out-of-plane field direction. Both the effective moment and Weiss temperature are comparable between *H*∥*ab* and *H*∥*c*.

## F. Absence of mixed valence

We present X-ray absorption data to directly confirm the $Eu^{2+}$ oxidation state without a mixed valence in $EuCd_2P_2$. This analysis is complementary to the magnetization data in Figs. 2 b,c in the text. The Eu *L$_3$*-edge X-ray absorption spectroscopy (XAS) involves $2p_{3/2} \rightarrow 5d$ excitations, and is very sensitive to the occupation of Eu *4f* electronic orbitals including presence of mixed or fluctuating valence. Absorption peaks ("white line") of $4f^7 5d^0$ ($Eu^{2+}$) and $4f^6 5d^0$ ($Eu^{3+}$) configurations are separated by approximately 8 eV [3]. The XAS data in Fig. S6 confirms the $4f^7$ (2+) valence of Eu ions in the $EuCd_2P_2$ compound without any detectable presence of a 3+ component at any temperature, both in zero field and *H*=2 T. This indicates that valence fluctuations are not involved in the mechanism driving the giant magnetoresistance in this material. Note that the internal clock of the XAS measurement, of about 50 attosec, is significantly faster than the characteristic time scale for valence fluctuations, $\tau \leq \hbar/\Delta E$ where $\Delta E$ is *4f* bandwidth, so XAS would detect the two valence components separately if valence fluctuations were present.

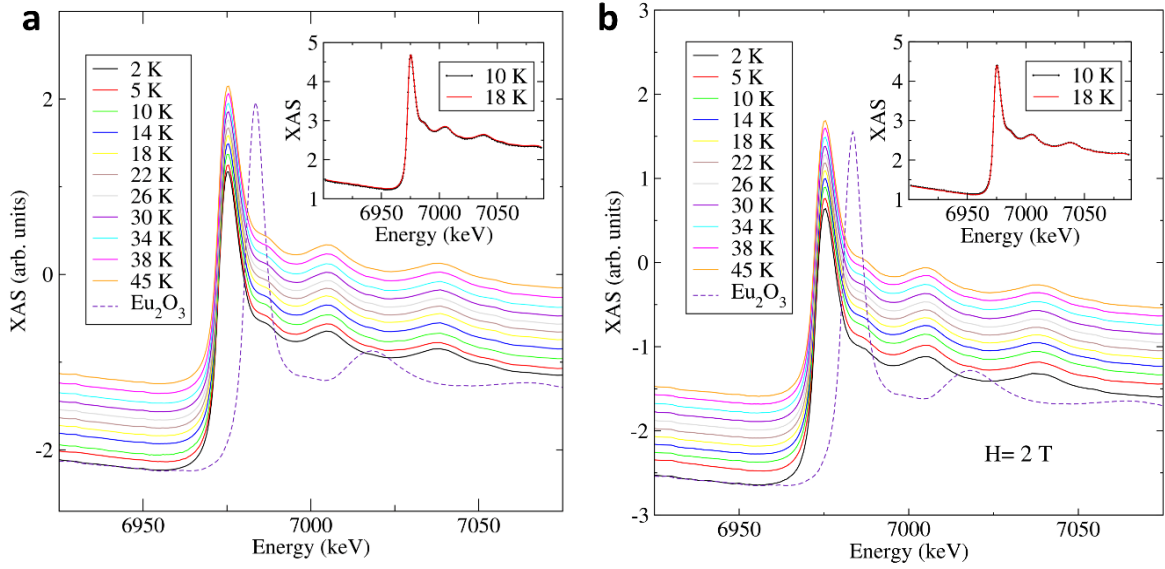

**Figure S6.** (a) XAS spectra at several temperatures below and above $T_N$ in zero field. The spectra are consistent with $Eu^{2+}$. Inset shows identical spectra at 10 and 18 K, just below and above $T_N$. The dashed line shows the spectrum of a $Eu^{3+}$ reference material $Eu_2O_3$. (b) Similar spectra but in a magnetic field of 2 T. The mixed valence is absent in both $H$=0 and 2 T.

### G. Absence of lattice distortion

In the main text (Figure 2d), we show the diffraction spectra for both (003) and ($\bar{1}$06) reflections without any analysis. It was clear from the raw data that the peaks remain unchanged as the temperature was varied through $T_N$. Here, we analyze those peaks shapes by fitting the spectrum at each temperature to a Voigt function (a sum of Gaussian and Lorentzian peak shapes) and subtracting a constant, temperature-dependent background. We extract the positions of the peaks from the fits and plot them in Figure S7 as a function of temperature. Due to considerable twining in the sample, we had to perform realignment procedures that introduced experimental error due to domain switching. As such, the error bars correspond to the largest difference between the adjacent fitting positions. The absence of anomalies in the temperature evolution of the peak positions in Figure S7a,b confirms the absence of Jahn-Teller-like lattice distortions in $EuCd_2P_2$.

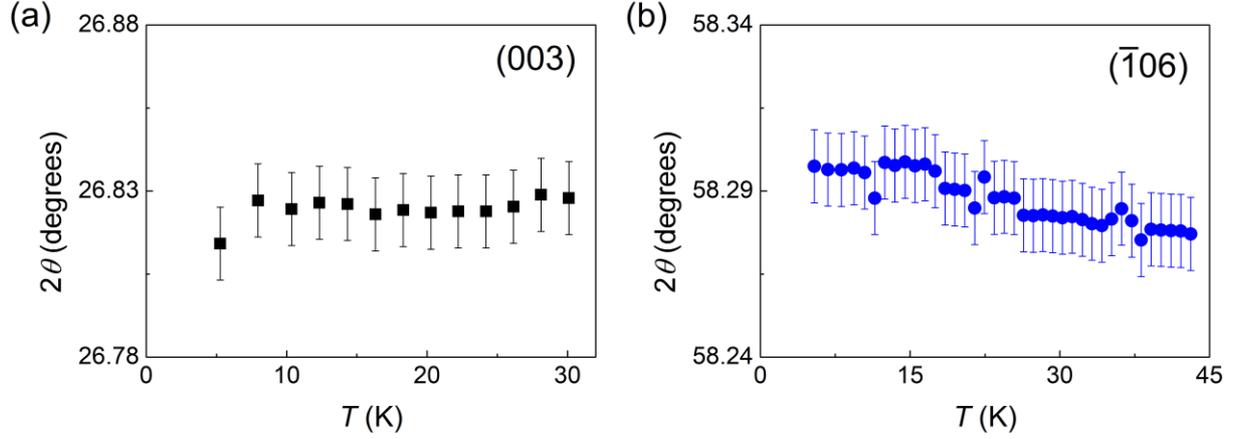

**Figure S7.** The Bragg peak positions ($2\theta$) are extracted from the synchrotron diffraction data and plotted as a function of temperature for both (003) and ($\bar{1}$06) reflections in panels (a) and (b), respectively. There are no anomalies at $T_N = 11$ K.

## H. Anisotropy of resistivity with respect to the field and current directions

As pointed out in the main text, CMR in $EuCd_2P_2$ is nearly independent of the direction of the magnetic field ($H_c$ vs. $H_{ab}$), but it depends strongly on the direction of the electric current ($\rho_c$ vs. $\rho_{ab}$). In Figure S8, we show four different configurations to measure CMR with current in-plane (top row) and out-of-plane (bottom row), and the field in-plane (left column) and out-of-plane (right column). Regardless of the field direction, the in-plane CMR ($J \| ab$) is of order $-10^3$ % and the out-of-plane CMR ($J \| c$) is of order $-10^4$ %. We show a 360° scan of $\rho_c$ in a few representative field values in Figure S9. The maximum anisotropy is a factor of 1.5 at 0.1 T, which is quite small and similar to $\rho_{ab}$ in the main text (Figure 1d).

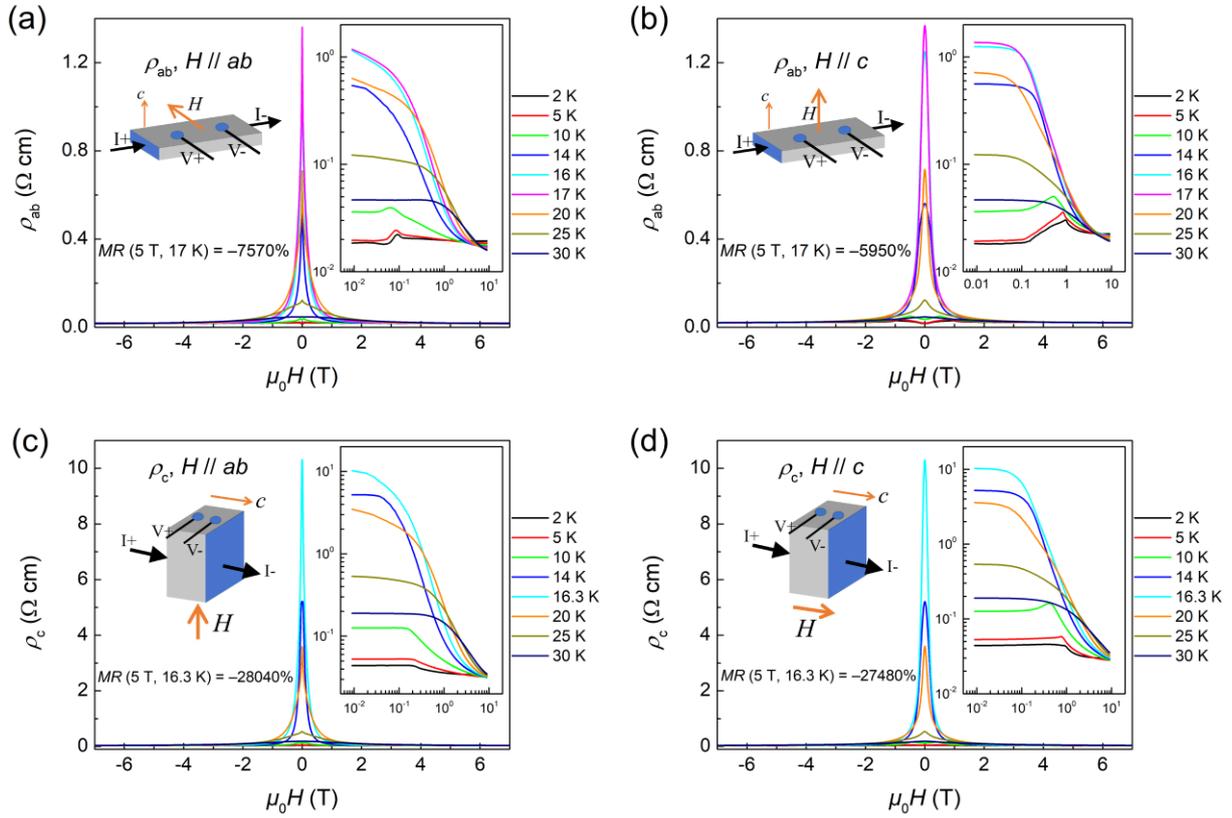

**Figure S8.** The field dependence of the electrical resistivity is shown with (a) in-plane current ($J\|ab$) and in-plane field ($H\|ab$), (b) in-plane current ($J\|ab$) and out-of-plane field ($H\|c$), (c) $J\|c$ and $H\|ab$, and (d) $J\|ab$ and $H\|c$. CMR depends mildly on the field direction, but strongly on the current direction.

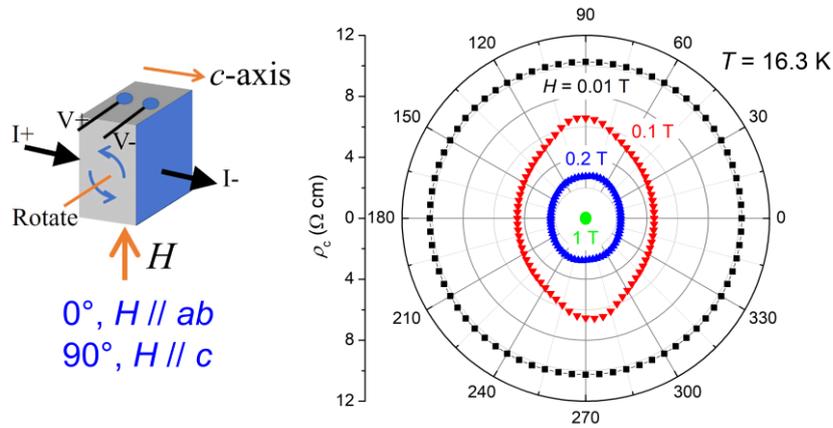

**Figure S9.** CMR has a weak dependence on the angle between the magnetic field and the electrical current, both when the current is out-of-plane (here) and in-plane (main Figure 1d).